\documentclass[11pt,a4paper]{amsart}
\usepackage[margin=25mm]{geometry}
\usepackage[numbers]{natbib}
\usepackage{hyperref}

\usepackage{amssymb,amsmath}
\usepackage{amsthm}
\usepackage{bbm, bm}
\usepackage{stmaryrd}
\usepackage{color}
\usepackage{graphicx}
\usepackage{caption}
\usepackage{float}
\usepackage{wrapfig}
\usepackage{subcaption}
\usepackage{amsaddr}

\usepackage{enumerate}
\usepackage{lineno}

\usepackage{abstract}

\input xy
\xyoption{all} 

\numberwithin{equation}{section}

\newtheorem{theorem}{Theorem}[section]

\theoremstyle{definition}

\newtheorem{example}[theorem]{Example}

 \newtheorem*{nogo}{No-Go Theorem}

\newcommand{\Diffeo}{\mathrel{\raisebox{-.25ex}{$\xrightarrow{\sim}$}}}

\newcommand{\rmd}{\textnormal{d}}

\newcommand{\phiA}{\boldsymbol{\phi}_\mathbb{A}}

\DeclareMathOperator{\Span}{Span}

\font\black=cmbx10 \font\sblack=cmbx7 \font\ssblack=cmbx5 \font\blackital=cmmib10  \skewchar\blackital='177
\font\sblackital=cmmib7 \skewchar\sblackital='177 \font\ssblackital=cmmib5 \skewchar\ssblackital='177
\font\sanss=cmss10 \font\ssanss=cmss8 
\font\sssanss=cmss8 scaled 600 \font\blackboard=msbm10 \font\sblackboard=msbm7 \font\ssblackboard=msbm5
\font\caligr=eusm10 \font\scaligr=eusm7 \font\sscaligr=eusm5  \font\fraktur=eufm10
\font\sfraktur=eufm7 \font\ssfraktur=eufm5 
\font\bsymb=cmsy10 scaled\magstep2
\def\all#1{\setbox0=\hbox{\lower1.5pt\hbox{\bsymb
       \char"38}}\setbox1=\hbox{$_{#1}$} \box0\lower2pt\box1\;}
\def\exi#1{\setbox0=\hbox{\lower1.5pt\hbox{\bsymb \char"39}}
       \setbox1=\hbox{$_{#1}$} \box0\lower2pt\box1\;}

\def\tx#1{{\fam0\relax#1}}

\newfam\bifam
\textfont\bifam=\blackital \scriptfont\bifam=\sblackital \scriptscriptfont\bifam=\ssblackital

\newfam\blfam
\textfont\blfam=\black \scriptfont\blfam=\sblack \scriptscriptfont\blfam=\ssblack

\newfam\bbfam
\textfont\bbfam=\blackboard \scriptfont\bbfam=\sblackboard \scriptscriptfont\bbfam=\ssblackboard

\newfam\ssfam
\textfont\ssfam=\sanss \scriptfont\ssfam=\ssanss \scriptscriptfont\ssfam=\sssanss
\def\sss#1{{\fam\ssfam\relax#1}}

\newfam\clfam
\textfont\clfam=\caligr \scriptfont\clfam=\scaligr \scriptscriptfont\clfam=\sscaligr

\newfam\frfam
\textfont\frfam=\fraktur \scriptfont\frfam=\sfraktur \scriptscriptfont\frfam=\ssfraktur

\def\hpb#1{\setbox0=\hbox{${#1}$}
    \copy0 \kern-\wd0 \kern.2pt \box0}
\def\vpb#1{\setbox0=\hbox{${#1}$}
    \copy0 \kern-\wd0 \raise.08pt \box0}

\def\pmb#1{\setbox0\hbox{${#1}$} \copy0 \kern-\wd0 \kern.2pt \box0}
\def\pmbb#1{\setbox0\hbox{${#1}$} \copy0 \kern-\wd0
      \kern.2pt \copy0 \kern-\wd0 \kern.2pt \box0}
\def\pmbbb#1{\setbox0\hbox{${#1}$} \copy0 \kern-\wd0
      \kern.2pt \copy0 \kern-\wd0 \kern.2pt
    \copy0 \kern-\wd0 \kern.2pt \box0}
\def\pmxb#1{\setbox0\hbox{${#1}$} \copy0 \kern-\wd0
      \kern.2pt \copy0 \kern-\wd0 \kern.2pt
      \copy0 \kern-\wd0 \kern.2pt \copy0 \kern-\wd0 \kern.2pt \box0}
\def\pmxbb#1{\setbox0\hbox{${#1}$} \copy0 \kern-\wd0 \kern.2pt
      \copy0 \kern-\wd0 \kern.2pt
      \copy0 \kern-\wd0 \kern.2pt \copy0 \kern-\wd0 \kern.2pt
      \copy0 \kern-\wd0 \kern.2pt \box0}

\mathchardef\za="710B  
\mathchardef\zb="710C  
\mathchardef\zg="710D  
\mathchardef\zd="710E  
\mathchardef\zve="710F 
\mathchardef\zz="7110  
\mathchardef\zh="7111  
\mathchardef\zvy="7112 
\mathchardef\zi="7113  
\mathchardef\zk="7114  
\mathchardef\zl="7115  
\mathchardef\zm="7116  
\mathchardef\zn="7117  
\mathchardef\zx="7118  
\mathchardef\zp="7119  
\mathchardef\zr="711A  
\mathchardef\zs="711B  
\mathchardef\zt="711C  
\mathchardef\zu="711D  
\mathchardef\zvf="711E 
\mathchardef\zq="711F  
\mathchardef\zc="7120  
\mathchardef\zw="7121  
\mathchardef\ze="7122  
\mathchardef\zy="7123  
\mathchardef\zf="7124  
\mathchardef\zvr="7125 
\mathchardef\zvs="7126 
\mathchardef\zf="7127  
\mathchardef\zG="7000  
\mathchardef\zD="7001  
\mathchardef\zY="7002  
\mathchardef\zL="7003  
\mathchardef\zX="7004  
\mathchardef\zP="7005  
\mathchardef\zS="7006  
\mathchardef\zU="7007  
\mathchardef\zF="7008  
\mathchardef\zW="700A  
\mathchardef\zC="7009  

\newcommand{\be}{\begin{equation}}
\newcommand{\ee}{\end{equation}}

\newcommand{\bea}{\begin{eqnarray}}
\newcommand{\eea}{\end{eqnarray}}
\def\*{{\textstyle *}}
\newcommand{\R}{{\mathbb R}}

\newcommand{\Z}{{\mathbb Z}}

\newcommand{\s}{{\textstyle *}}

\def\Sec{\sss{Sec}}
\def\Sol{\sss{Sol}}

\def\sT{{\sss T}}

\def\xi{\tx{i}}

\def\s*{{\scriptstyle *}}

\newcommand{\beas}{\begin{eqnarray*}}
\newcommand{\eeas}{\end{eqnarray*}}

\title{Frozen Motion: Why Single Carrollian Scalars Cannot Propagate} 
\author{Andrew James Bruce 
 }  
\address{Independent Researcher, Cardiff, United Kingdom }
   \email{andrewjamesbruce@googlemail.com}

   \date{\today}
 
\begin{document}
 \maketitle
\vspace{-20pt}
\begin{abstract}{\noindent We investigate a class of first-order scalar field theories minimally coupled to a Carrollian connection that are defined intrinsically on the Carrollian plane, i.e., the theories are not defined via limits of Lorentzian theories.  The theories built are invariant under the extended Carrollian transformations which include supertranslations.  The symmetry allows for a large class of Lagrangians, independence of spacetime coordinates is all that is required. However, invariance under supertranslations (which include boosts as linear supertranslations) forces the energy density to be static and the momentum density to vanish -- this precludes on-shell propagation of fields.  Thus, to have propagating theories, one must move beyond single field theories that are minimally coupled to the geometry.  }\\
\noindent {\Small \textbf{Keywords:} Carrollian Field Theory;~Symmetries \& Conservation Laws}\\
\noindent {\small \textbf{MSC 2020:} \emph{Primary:}~70G45~\emph{Secondary:}~70S10;~81R10}
\end{abstract}
\medskip
\begin{flushright}
\emph{``But I don't want to go among mad people,'' Alice remarked. }\\
Lewis Carroll, Alice's Adventures in Wonderland,  (1865) 
\end{flushright}
\section{Introduction}
The Carrollian group, understood as the $c\rightarrow 0$ limit of the Poincaré group, was first explored by Lévy-Leblond \cite{Lévy-Leblond:1965} and Sen Gupta \cite{SenGupta:1966} in the mid-1960s (also see Henneaux \cite{Henneaux:1979}). In this limit, light cones collapse into lines, and massive particles become ultra-local, i.e., they are frozen in space and can only move through time.  In other words, all spacetime events at different spatial points become causally disconnected.  Amazingly, this limit has attracted considerable attention over the past decade from various perspectives, including flat-space holography, hydrodynamics, and condensed matter physics; for a review, the reader may consult Bagchi et al. \cite{Bagchi:2025}. While Carrollian physics is usually understood in terms of the vanishing-speed-of-light limit, there is a drive to understand physics in an intrinsic framework.  In particular, Carrollian scalar field theories are typically either electric or magnetic, i.e., the dynamics are governed by temporal derivatives or spatial derivatives. An intrinsic formulation allows for theories that are neither electric nor magnetic, see \cite{Bagchi:2023,Ciambelli:2024,Ecker:2024,Saha:2022,Tadros:2024}. We take the position that if the Carrollian group is to be taken as a fundamental part of physics, then field theories that are invariant under Carrollian boosts need to be constructed and studied, even if this leads to strange or seemingly paradoxical behaviour when compared to Poincaré invariant theories.\par 
In this note, we geometrically build (local) single scalar field theories on the Carrollian plane $(\R^2, g, \kappa)$, where, in global coordinates $(t,x)$,  $g = \rmd x \otimes \rmd x$, and $\kappa = \partial_t$ (see Duval et al. \cite{Duval:2014a,Duval:2014b,Duval:2014c} for the notion of a Carrollian manifold). The infinite group of isometries of the degenerate metric contains the standard $1+1$-dimensional Carroll transformations as well as supertranslations.  That is, we consider transformations (coordinate changes) of the form 
$$t' = t - \alpha - \beta\, x -  \gamma f(x)\,, \qquad x' = x - \delta \,,$$
which we refer to as the extended Carroll transformations. The standard Carroll transformations are defined by setting $\gamma =0$. Supertranslations are understood as $\gamma \neq 0$ and  non-zero $f(x) \neq x$. We stress these transformations do not constitute the full conformal Carrollian group $\mathrm{CCarr}_2 \cong \mathrm{BMS}_3$, but rather we consider a subgroup thereof that preserves the exact form of the degenerate metric.  In particular, we drop superrotations, which, written infinitesimally, are of the form
$$t' = t - (n+1)\, \epsilon_n \, t\, x^n\,, \qquad x' = x - \epsilon_n \, x^{n+1}\,, $$
with $n \in \Z$. See \cite{Duval:2014a,Duval:2014b} for further details, including the isomorphism between conformal Carrollian groups and BMS groups.\par
We remark that the Bondi--van der Burg--Metzner--Sachs (BMS) transformations are vital in general relativity, and represent the symmetry group of asymptotically flat spacetimes at null infinity. For a review of the BMS symmetries, the reader may consult Kervyn \cite{Kervyn:2025}.\par 
We employ techniques from the theory of jet bundles to establish a general form of scalar field theories invariant under the above transformations (for an introduction to jet bundles, the reader may consult Sardanashvily \cite[Chapter 2.1]{Sardanashvily:2013}, and for a comprehensive review the reader may consult Saunders \cite{Saunders:1989}).  Importantly, we build the theory from the `ground up', assuming as little as possible; and we do not use any deep theorems/construction from the general theory of jet bundles.  Note that the Carrollian plane is a principal $\R$-bundle where the temporal direction defines the fibres. The $\R$-action is essentially a shift in $t$, and the extended Carroll transformations respect this action. The partial derivative $\partial_x$ is not invariant under the extended Carroll transformations, and so we require a clock form $\tau$, equivalently an Ehresmann connection, to invariantly define vertical and horizontal vector fields. We will refer to such an Ehresmann connection as a Carrollian connection, and remark that it need not be a principal connection in general. In the jet bundle formalism, a Carrollian connection is understood as a decomposition (a choice of a class of coordinates) of the first jet bundle that renders the question of constructing invariant actions clear to answer. In particular, via a Carrollian connection we construct jet coordinates that are scalar under the extended Carroll transformations.  The theories constructed are scalar theories minimally coupled to the Carrollian connection, so they do not exhaust all possible scalar field theories.  \par
We comment that the Carrollian connection (or clock form), from the perspective of this note, is a fixed gauge field introduced to remedy the non-covariance of the spatial derivative under the extended Carroll transformations. This is quite distinct from introducing an affine connection where the notion of parallel transport is the motivation. We will not consider affine connections in this note and direct the interested reader to \cite{Bruce:2026a} for details of affine connection on Carrollian manifolds within the framework of Lie algebroids.   \par
The possible Lagrangians are almost unconstrained; we only require that they do not explicitly depend on spacetime coordinates; the form of the Lagrangian may be very different to those found in Lorentzian theories. However, invariance under Carrollian boosts, or more generally, supertranslations, places stringent constraints on the solution of the equation of motion; specifically, the energy density must be static, and the momentum density must identically vanish.  These conditions preclude propagation of the fields and render the theories ultra-local.  While this result is not unexpected from the point of view of the electric and magnetic Carrollian limits of Lorentzian theories, the core mechanism examined in this note is the geometric symmetry of the Lagrangians. Thus, the difficulty in constructing propagating Carrollian theories for a single minimally coupled scalar is not a matter of finding a sufficiently exotic Lagrangian.  Of course, the results of this note do not rule out propagating Carrollian theories; however, such theories cannot be as minimal as those presented here.\par 
We remark  Ecker et al. \cite{Ecker:2024} constructed multi-scalar field theories that allow for propagation; these kinds of theories are under the umbrella of swifton theories, which are similar to tachyon theories but are energetically stable. As such theories involve more than one field and careful couplings between them, they fall outside our minimalistic framework. Tadros  \& Kolář  \cite{Tadros:2024} have constructed higher-derivative Carrollian scalar field theories.  We further comment that systems of Carrollian particles may show relative motion under suitable conditions, see  Bergshoeff et al. \cite{Bergshoeff:2014}, Casalbuoni et al. \cite{Casalbuoni:2023}, and Zhang et al. \cite{Zhang:2024}.  In $2+1$ dimensions, the Carroll group admits a two-parameter central extension, and a particle may exhibit non-trivial dynamics (unlike free Carroll particles in higher dimensions) due to the two additional Casimir invariants, see Marsot \cite{Marsot:2022}, Marsot et al. \cite{Marsot:2023}, and Zeng et al. \cite{Zeng:2025}. 
%
%
\section{Carrollian Geometry and the Models}
\subsection{The Canonical  Carrollian Geometry of the Plane}
The manifold that we will build the scalar field theories on is $M \cong \R^2$, which we equip with standard global coordinates $(t,x)$. The canonical (weak) Carrollian structure on $M$ is given by $g = \rmd x \otimes \rmd x$ and $\kappa = \partial_t$; clearly $\ker(g) = \Span \{\kappa \}$.  The admissible changes of coordinates we consider are the \emph{extended Carroll transformations} (see \cite{Lévy-Leblond:1965})
\begin{align}\label{eqn:ConCarTran}
& t' = t - \alpha - \beta\, x -  \gamma f(x)\,, && x' = x - \delta \,,
\end{align}
where $\alpha, \beta, \gamma, \delta$ are constants (with the appropriate units,) and $f \in C^\infty(\R)$ is an arbitrary smooth function.  The specific transformations $t' = t - \gamma\, f(x)$ are referred to as \emph{supertranslations}, and shows that the isometry group here is infinite dimensional; the extension we consider is BMS-like.\par 
We observe that 
$$\rmd t' = \rmd t -  \rmd x\, \big( \beta + \gamma \, \partial_x f(x) \big) \,, \qquad \rmd x' = \rmd x\,,$$
and thus the (degenerate) metric is preserved under \eqref{eqn:ConCarTran}. A direct calculation gives 
\begin{equation}\label{eqn:ParDerChange}
\frac{\partial}{\partial t'} = \frac{\partial}{\partial t}\,, \qquad  \frac{\partial}{\partial x'}  = \frac{\partial}{\partial x}+ \big( \beta + \gamma \,\partial_x f\big)\frac{\partial}{\partial t}\,,
\end{equation}
meaning that $\kappa = \partial_t$ is invariant under the extended Carroll transformations.  In short, the Carrollian structure $(g, \kappa)$ is preserved under \eqref{eqn:ConCarTran}. \par 
Observe that we have a principal $\R$-bundle $\mathrm{prj}: M \rightarrow \Sigma\cong \R$, where the right action in coordinates is  $(t, x)\triangleleft r = (t+r, x)$. The vertical transformation part of the extended Carroll transformations \eqref{eqn:ConCarTran} we thus interpret as gauge transformations. Moreover, the degenerate metric is right invariant, or in more physical language, the metric is static. \par
We will need a volume on $M$ in order to build an action. We will choose the coordinate volume $\rmd t \wedge \rmd x$. To check if this volume is invariant under the extended Carroll transformations, we examine the Jacobian
\begin{equation}\label{eqnJac}
\mathrm{Jac} =  \det \begin{pmatrix}
    \frac{\partial t'}{\partial t} & \frac{\partial t'}{\partial x} \\
     \frac{\partial x'}{\partial t} & \frac{\partial x'}{\partial x}
\end{pmatrix} = \det \begin{pmatrix}
   1 & \star \\
    0 & 1
\end{pmatrix} = 1\,.
\end{equation}
%
%
\subsection{Jet Bundle Geometry and  Carrollian Connections}
For background material on jets and connections, the reader may consult Sardanashvily \cite[Chapters 1, 2, 3]{Sardanashvily:2013} or Kolář et al.  \cite[Chapters III, IV]{Kolář:1993}. \par 
We will consider scalar fields to be sections of the Thomas bundle\footnote{The construction is due to T.Y. Thomas in the early 1920's in relation to projective geometry (see \cite{Thomas:1925}).} $\pi : \hat{M} \rightarrow M$, which is, due to $M \cong \R^2$ being contractible, a trivial line bundle.  We equip $\hat{M}$ with adapted coordinates $(t, x, y)$, which under general admissible coordinate changes transform as 
$$t' = t'(t,x)\,, \qquad x' = x'(t,x)\,, \qquad y ' = y \, |\mathrm{Jac}(t,x)|^{-1}\,,$$
where $\mathrm{Jac}$ is the Jacobian of the spacetime coordinate changes.  Observe that under the extended Carroll transformations \eqref{eqn:ConCarTran}, via evaluating the Jacobian (see \eqref{eqnJac}), the fibre coordinate of the Thomas bundle remains unchanged.  From this point on, we will only consider the extended Carroll transformations. Sections $\phi \in \Sec(\hat{M})$ are identified with (real) scalar fields, and we will write $\phi^* y := \phi(t,x)$.\par 
The (first-order) jet of a section at $m \in M$, $j_m \phi$ is an equivalence class of sections $\phi \in \Sec(\hat{M})$ that are identified by their Taylor series at $m = (t_0, x_0)$ to first order.  That is, $\tilde{\phi} \in j_m \phi$ if 
$$\tilde{\phi}(t_0, x_0) = \phi(t_0, x_0)\, \qquad \partial_t \widetilde{\phi}(t_0, x_0) =\partial_t \phi(t_0, x_0)\,, \qquad \partial_x \widetilde{\phi}(t_0, x_0) = \partial_x \phi(t_0, x_0)\,.$$
The (first) jet bundle is defined as $J\hat{M} := \bigcup_{m \in M} j_m \phi$, which inherits the structure of a smooth bundle from $M$. We denote the fibre bundle as $J \phi : J \hat{M} \rightarrow M$, and employ adapted coordinates  $(t,x, y, y_t, y_x)$, where the admissible coordinate transformations are the extended Carrollian transformation \eqref{eqn:ConCarTran} and their induced consequences 
\begin{equation}
y'= y\,, \qquad  y'_t = y_t\,, \qquad y'_x = y_x + \big(\beta + \gamma \, \partial_xf(x)\big)\, y_t\,.
\end{equation}
Notice that the coordinate $y_x$ is not invariant under the admissible coordinate changes.  To correct for this, we introduce a decomposition (over $M$) $\Gamma : J\hat{M} \Diffeo \hat{M} \times \R \times \R$. By employing coordinates $(t,x, y, y_t, \hat{y}_x)$ the diffeomorphism $\Gamma$ is specified by 
\begin{equation}\label{eqn:Connection}
\Gamma^*(\hat{y}_x) := y_x + \mathbb{A}_x^{~t}(t,x) \, y_t\,,
\end{equation}
where $\mathbb{A}^{'~~t}_x = \mathbb{A}_x^{~t} - \partial_xt'   =  \mathbb{A}_x^{~t} -\big(\beta + \gamma \, \partial_xf(x)\big)$ under the admissible changes of coordinates. We recognise we have an Ehresmann connection\footnote{ Recall that an Ehresmann connection on a fibre bundle $E$ is the assignment of a (smooth) subbundle $\mathsf{H}E $ of $\sT E$  that is complimentary to the vertical subbundle $\mathsf{V}E$, i.e., $\sT E = \mathsf{V}E \oplus \mathsf{H}E$. Here we describe this locally using coordinates. For further details, the reader may consult \cite[Chapter III]{Kolář:1993}. } on the principal $\R$-bundle $\mathrm{prj}: M \rightarrow \Sigma$. Thus, from the general theory of jets and connections, we have a section of the jet bundle $ JM \rightarrow M$. Moreover, from the general theory, we know such sections always exist (see, for example, \cite[Lemma 11.3]{Kolář:1993}). However, the connection is not unique, nor is there a canonical choice. We will refer to $\mathbb{A}$ as the \emph{Carrollian connection}.  We understand the Carrollian connection as a fundamental part of the background geometry; we do not consider it a dynamical field in the following.\par 
The invariant coframe associated with a Carrollian connection is given by $\tau := \rmd t - \rmd x \, \mathbb{A}_x^{~t}$ and $\rmd x$. The one-form $\tau$ is commonly referred to as the \emph{clock form}. As $M$ is two-dimensional, the clock form satisfies the Frobenius condition for integrability, i.e., $\tau \wedge \rmd \tau =0$. The associated foliation, defined by the kernel of the clock form, splits $M$ into spatial leaves/cuts which define the notion of absolute simultaneity.  Note that the \emph{torsion of clock form} is locally given by $\rmd \tau = - \tau \wedge \rmd x \, \partial_t \mathbb{A}_x^{~t}$. Defining  the spatial covariant derivative as $\nabla_x := \partial_x + \mathbb{A}_x^{~t}(t, x) \partial_t$, we observe that $[\partial_t , \nabla_x] = (\partial_t \mathbb{A}_x^{~t})\, \partial_t=0$ if the torsion of the clock form vanishes. In other words, the Carrollian connection is flat if and only if the clock form is closed. As this foliation will play no explicit role in the following, we will not elaborate further and direct the reader to the original literature \cite{Duval:2014a,Duval:2014b,Duval:2014c}. 
%
%
\subsection{Building Lagrangians and Actions}
A Lagrangian density for a (single, real) scalar field on $(M = \R^2, g = \rmd x \otimes \rmd x, \kappa= \partial_t)$ is a horizontal two-form on $J\hat{M}$, i.e.,
$$\mathcal{L} \in \Omega^{2,0}\big(J\hat{M})\,.$$
Once a Carrollian connection $\mathbb{A}$ has been chosen, we can consider  the covariant or gauged Lagrangian 
$$\mathcal{L}_{\Gamma} := \big(\Gamma^{-1} \big)^* \mathcal{L} \in \Omega^{2,0}\big(\Gamma(J\hat{M})\big)\,. $$
We then observe that a covariant Lagrangian $\mathcal{L}_{\Gamma}$ is invariant under the extended Carroll transformations \eqref{eqn:ConCarTran}, if, in adapted coordinates, it is of the form
\begin{equation}\label{eqn:InvLag}
\mathcal{L}_{\Gamma} = \tau \wedge \rmd x\, L(y, y_t, \hat{y}_x)\,,
\end{equation}
where we have employed the clock form; note that $\tau \wedge \rmd x = \rmd t \wedge \rmd x$. This follows as the coordinate volume is invariant under the extended Carroll transformations and the admissible coordinate transformation on the split jet bundle. It is important to note that we have constructed minimally coupled Lagrangians; the Carrollian connection only appears via the decomposition of the jet bundle. Other than being minimally coupled, the Lagrangians are general, and not required to be non-degenerate nor of mechanical-type.\par 
To construct actions, let us suppose that we fix a diffeomorphism $\Gamma$, i.e., we have chosen a Carrollian connection $\mathbb{A}$. Any section $\phi \in \Sec(\hat{M})$ can be prolongated to a section $j_{\mathbb{A}}\phi \in \Sec(\Gamma(J\hat{M}))$, via taking derivatives. In fibre coordinates, we define 
$$\big(j_{\mathbb{A}}\phi\big)^*(y, y_t, \hat{y}_x)= \big(\phi(t,x), \,\partial_t \phi(t,x),\, \nabla_x \phi(t,x) \big) =: \phiA \,,$$
where we have define the spatial covariant derivative $\nabla_x := \partial_x + \mathbb{A}_x^{~t}(t, x) \partial_t$. Actions are then defined as 
\begin{equation}
\mathsf{S}[\phi] = \int_M \big( j_{\mathbb{A}}\phi\big)^* \mathcal{L}_\Gamma\,,
\end{equation}
where we insist that $\big( j_{\mathbb{A}}\phi\big)^* L \in L^1(\R^2)$.  For example, we  may restrict to compactly supported smooth sections to ensure the action is well-defined and force boundary terms to vanish. Note, we are considering actions that are local functionals, as standard in classical field theory.\par 
The generalised Euler--Lagrange equations are of the standard form in the presence of a covariant derivative, i.e., 
\begin{equation}\label{eqn:GELeq}
 \mathbb{EL}(\boldsymbol{\phi}_\mathbb{A})  =   \dfrac{\partial }{\partial \phi}L(\phiA) - \partial_t\left( \frac{\partial L(\phiA)}{\partial (\partial_t \phi)}\right) - \nabla_x \left( \frac{\partial L(\phiA) }{\partial (\nabla_x \phi)}\right) =0\,.
\end{equation}
A section $\phi \in \Sec(\hat{M})$ is said to be a \emph{solution} of the theory  if it satisfies 
\begin{enumerate}[(i)]
\item $\big( j_{\mathbb{A}}\phi\big)^* L \in L^1(\R^2)$; and
\item $\mathbb{EL}(\boldsymbol{\phi}_\mathbb{A}) =0$.
\end{enumerate}
The first condition says the action is well-defined, and the second condition says that the generalised Euler-- Lagrange equations hold. The set of solutions we will denote as $\Sol_{\mathcal{L}_\Gamma}(\hat{M})$.
\begin{example}[Electric-type]
As an example of an electrical-type Lagrangian, consider $L = \frac{1}{2} y^2_t - V(y)$. Evaluating the generalised Euler--Lagrange equation on the prolongation of a section produces  $\partial^2_t \phi + V'(\phi)=0$. Note that the equation of motion does not depend on the chosen Carrollian connection.
\end{example}
\begin{example}[Magnetic-type]
As an example of a magnetic-type Lagrangian, consider $L = \frac{1}{2} \hat{y}^2_x - V(y)$. Evaluating the generalised Euler--Lagrange equation on the prolongation of a section produces  $\nabla^2_x \phi + F\, \nabla_x\phi + V'(\phi)=0$, where the curvature is defined as $F := \partial_t \mathbb{A}_x^{~t}$. 
\end{example}
\begin{example}[Mixed-type]
As an example of a mixed-type Lagrangian, consider $L = \frac{1}{2} y^2_t-\frac{1}{2} \hat{y}^2_x - V(y)$. Evaluating the generalised Euler--Lagrange equation on the prolongation of a section produces  $\partial^2_t \phi -\nabla^2_x \phi - F\, \nabla_x\phi + V'(\phi)=0$, where the curvature is defined as $F := \partial_t \mathbb{A}_x^{~t}$. 
\end{example}
%
%
\subsection{Energy, Momentum, and Ultra-Locality}
The Noether one-form\footnote{ In $n+1$-dimensions, we have $\mathsf{J} :=  \textrm{Vol}_\Sigma  \, \mathcal{E} - \tau \wedge \boldsymbol{\mathcal{P}} \in \Omega^{n,0}(\Gamma(J \hat{M}))$, where $\boldsymbol{\mathcal{P}}$ is a horizontal $n-1$-form. The minus sign has been chosen as a convention to match standard constructions in Lorentzian theories. Note that  $\mathsf{J}$ mimics the role of a Hodge-dualised current in Lorentzian geometry. However, as there is no canonical Hodge dual here due to the degenerate nature of the metric, we are somewhat forced to work with a horizontal $n-1$-form from the start.  Fecko (see \cite{Fecko:2024}) has proposed an analogue of the Hodge operator on Galilean/Carrollian spacetimes. We direct the interested reader to Fecko for details. }  associated with temporal shifts $t' = t - \alpha$ can directly calculated
\begin{equation}\label{eqn:NoetherForm}
\mathsf{J} = \rmd x \, \mathcal{E} - \tau \, \mathcal{P} = \rmd x \left( y_t \frac{\partial L}{\partial y_t } +  \hat{y}_x \frac{\partial L}{\partial \hat{y}_x} - L \right) - \tau \, \left ( y_t \frac{\partial L}{\partial \hat{y}_x} \right )\in \Omega^{1,0}(\Gamma(J \hat{M}))\,.
\end{equation}
As standard,  we will refer to the components  $(\mathcal{E}, \mathcal{P})$ as the \emph{energy density}  and \emph{ momentum density}, respectively.  Note that these components are invariant under the extended Carrollian transformations, and are not the canonical energy and momentum density, as we have employed the invariant coframe defined by the clock form. However, these densities are considered physical observables as they are invariant (once a clock-form is given). This is quite different to Lorentzian field theories where energy and momentum are frame-dependent. 
\begin{example}
Consider $L = \frac{1}{2} y^2_t-\frac{1}{2} \hat{y}^2_x$, then a simple calculation shows that 
$$\mathsf{J} = \rmd x\, \left(\frac{1}{2} y^2_t-\frac{1}{2} \hat{y}^2_x \right) + \tau\, \big(y_t \hat{y}_x\big)\,.$$
In particular, $\mathcal{E} = L$ and $\mathcal{P} =-  y_t \hat{y}_x$. Let us compare this with the situation of a scalar field on $1+1$-dimensional Minkowski spacetime. The free action is $L_{mink} = \frac{1}{2} q_t^2 - \frac{1}{2}q_x^2$, which is invariant under the Lorentz transformations. The canonical energy and momentum densities are  (via a textbook calculation)
$$T^0_{~0} = \frac{1}{2} q_t^2 + \frac{1}{2}q_x^2\,, \qquad T^0_{~1} = - q_t q_x\,,$$ 
respectively. Formally, there is little difference here in the functional dependencies. However, the canonical components $T^0_{~\mu}$ in Minkowski spacetime are coordinate-dependent and transform non-trivially under Lorentz boosts. In contrast, because $\mathcal{E}$ and $\mathcal{P}$ are defined using the clock form $\tau$ and spatial coframe $\rmd x$, they are invariant under the extended Carroll transformations.
\end{example} 
The continuity equation for a solution $\phi \in \Sol_{\mathcal{L}_\Gamma}(\hat{M})$ is the closure of the Noether one-form, i.e., $\rmd\big((j_\mathbb{A}\phi)^* \mathsf{J})=0$, explicitly 
\begin{equation}\label{eqn:ContEQ}
\tau \wedge \rmd x \, \left ( \partial_t \big( (j_\mathbb{A}\phi)^*\mathcal{E} \big)+ F \, (j_\mathbb{A}\phi)^*\mathcal{P} + \nabla_x \big((j_\mathbb{A}\phi)^*\mathcal{P}  \big)\right)=0\,.
\end{equation} 
We can repeat the analysis for supertranslations $t' = t - \gamma \, f(x)$; we need only multiply the expression by the smooth function $f \in C^\infty(\R)$.  In particular, the relevant continuity equation is
 $$\rmd\big(f\, (j_\mathbb{A}\phi)^* \mathsf{J}) = \tau \wedge \rmd f \, (j_\mathbb{A}\phi)^*\mathcal{P} =0\, .$$
The previous continuity equation must hold for all smooth functions $f \in C^\infty(\R)$, and thus $(j_\mathbb{A}\phi)^*\mathcal{P} =0$ for all $\phi \in \Sol_{\mathcal{L}_\Gamma}(\hat{M})$. We then observe that the continuity equation \eqref{eqn:ContEQ} associated with temporal translations reduces to $\partial_t \big( (j_\mathbb{A}\phi)^*\mathcal{E} \big)=0$. That is, the extended Carrollian transformations force  the energy density to be static and the momentum density to vanish. The conserved current associated with a supertranslation is defined as $\mathsf{J}_f(\boldsymbol{\phi}_\mathbb{A}):=\big( j_\mathbb{A}\phi)^*\mathsf{J}\, f$, with $\phi \in \Sol_{\mathcal{L}_\Gamma}(\hat{M})$.\par
Note that the above analysis is independent of the chosen Carrollian connection/clock form. It must be stressed that Carrollian boosts are examples of linear supertranslations, and so the analysis holds true for $t' = t - \beta \, x$.  We need not consider the extended Carrollian transformations to reach the same conclusion. \par 
Examining the Noether one-form \eqref{eqn:NoetherForm}, then for $(j_\mathbb{A}\phi)^*\mathcal{P} =0$ for all solutions $\phi \in \Sol_{\mathcal{L}_\Gamma}(\hat{M})$ forces one of two conditions:
\begin{enumerate}[(i)]
\item For all $\phi \in \Sol_{\mathcal{L}_\Gamma}(\hat{M})$, $\partial_t \phi =0$; or \label{con:Mag}
\item The Lagrangian must be independent of $\hat{y}_x$, i.e., $L \neq L( \hat{y}_x)$. \label{con:Elec}
\end{enumerate}
The constraint \eqref{con:Mag} we refer to as the \emph{magnetic or static constraint}; while the Lagrangian may contain both temporal and spatial derivatives,  and so the equation of motion may contain the curvature, solutions are forced by the symmetries to be static.   The condition \eqref{con:Elec} on the action/Lagrangian leads to what we will refer to as \emph{theories of electric-type}. The conclusion is that the dynamics are ``frozen'' in these theories. \par 
We further more observe that, for static solutions $\partial_t\phi =0$, the action `splits'
$$\mathsf{S}[\phi] = \int_{-\infty}^{+\infty} \rmd t ~ \int_{-\infty}^{+\infty} \rmd x\, (j_\mathbb{A} \phi)^*L\,, $$
and for this to be well-defined, it must be the case that $\int_{-\infty}^{+\infty} \rmd x\, (j_\mathbb{A} \phi)^*L =0$. \par 
\medskip
\begin{small}
\noindent \textbf{Aside:} If instead of fixing the Carrollian connection we treat it as dynamical and vary the action, we observe that 
$$ \dfrac{\delta \mathsf{S}[\phi, \mathbb{A}]}{\delta \mathbb{A}_x^{~t}}=0 \Longrightarrow\frac{\partial L(\boldsymbol{\phi}_\mathbb{A})}{\partial \mathbb{A}_x^{~t}} = \frac{\partial \nabla_x\phi}{\partial \mathbb{A}_x^{~t}} \frac{\partial L(\boldsymbol{\phi}_\mathbb{A})}{\partial( \nabla_x\phi)}  =  \partial_t \phi \frac{\partial L(\boldsymbol{\phi}_\mathbb{A})}{\partial( \nabla_x\phi)} = (j_\mathbb{A}\phi)^*\mathcal{P} =0\,.$$
Thus, we recover the condition that the momentum density identically vanishes on-shell.
\end{small}
\begin{example}[Magnetic Non-Mechanical] Consider the Lagrangian $L = A \sin(\hat{y}_x/B)$,  which has been chosen to illustrate the constructions rather than with physical applications in mind. The associated generalised Euler--Lagrange equation,
$$\sin (\partial_x \phi/B) \, \partial_x^2 \phi  + F \, B \cos(\partial_x \phi/B) =0\,,$$
where we have imposed the static constraint $\partial_t \phi =0$. Observe that $L(\boldsymbol{\phi}_\mathbb{A})=0$ for configurations of the form $\phi_n(x) := 2 \pi Bn \, x + \phi_0$, with $n \in \Z$.  Importantly, this implies $\mathsf{S}[\phi_n] = 0$, and thus the action is well-defined. Examining the generalised Euler--Lagrange equation,  we encounter two conditions:
\begin{enumerate}[(a)]
\item $F \neq 0$ forces the only solution to be $n =0$, and $(j_\mathbb{A}\phi_0)^* \mathcal{E} = 0$; or\label{cond:Fneq0}
\item $F =0$,  solutions exist for all $n \in \Z$, and  $(j_\mathbb{A}\phi_n)^* \mathcal{E} =  2 \pi A \, n$.  \label{cond:Feq0}
\end{enumerate}
In particular, \eqref{cond:Feq0} allows for a static but non-zero energy density. 
\end{example}
\begin{small}
\noindent \textbf{Aside:} Moving past minimal coupling, we can consider Lagrangians of the 
form  $L = L(\phi, \partial_t\phi , \nabla_x \phi, F)$, that is we can include torsion terms.  The Noether one-form $\mathsf{J}$ associated with temporal shifts is formally unchanged; the exact form of $\mathcal{E}$ and $\mathcal{P}$ will, in general, be a little more involved. Nonetheless, the continuity equation takes the same form, and so invariance under supertranslations imposes, notationally ignoring the evaluation on a section,  $\tau \wedge \rmd f \,  \mathcal{P} =0$, thus on-shell $\mathcal{P}=0$.  And so $\partial_t \mathcal{E}=0$ even in the presence of torsion terms. 
\end{small}
%
%
\section{Concluding Remarks}
In this note, we have examined the classical theory of a single (real) scalar minimally coupled to a Carrollian connection on the Carrollian plane that is invariant under the extended Carroll transformation, i.e., the standard Carroll transformations together with supertranslations.  The theory of jet bundles was applied to simplify the study. We stress that the theories built are intrinsic and are not  build using the ultra-relativistic limit $c\rightarrow 0$. The core results may be stated as follows.
\begin{enumerate}[i)]
\item  The form of the Lagrangians permitted is quite general; they cannot explicitly depend on spacetime coordinates $(t,x)$.
\item Supertranslations $t' = t - \gamma\, f(x)$ (including restricting to Carrollian boosts) either force solutions of the (generalised) Euler-Lagrange equations to be static, or the Lagrangian must be independent of the spatial covariant derivative of the fields.  In essence, the electric/magnetic split remains in these intrinsic theories and is not a unique feature of the Carrollian limit.  
\end{enumerate}
Interpreting these results, we see that defining a minimally coupled single real scalar field theory action on the Carrollian plane that has a (quasi-)hyperbolic equation of motion is trivial; an example is $L \sim y^2_t - \hat{y}^2_x$. However, the solutions are forced by the symmetry to be static. Thus,  such theories do not allow (on-shell) propagation of small disturbances. \par 
It is well-known that in $1+1$-dimensions, the Carrollian group and  Galilean groups are isomorphic, i.e., $\mathrm{Carr}_2 \cong \mathrm{Gal}_2$.  Geometrically, the duality is realised by $ t \leftrightarrow  x$.  Thus, the core result of this note dualises; one cannot build first-order single (real) scalar field theories minimally coupled to a Newton--Cartan connection on the plane $\R^2$ that are invariant under Galilean boosts $x' = x - v t$, and admit propagating solutions.  In part, this provides a geometric rationale for why the (time-dependent) Schrödinger equation--the standard dynamical Galilean field theory--necessarily requires complex wavefunctions, which can be viewed as a pair of real fields. The main result of this note sits comfortably with Galilean scalar field theories. In particular,  first-order single scalar field theories that are Galilean invariant do not allow propagation.   This should be contrasted with Galileon field theories, where higher-order derivatives of the fields appear in the Lagrangian; such theories do allow propagation (see Nicolis et al. for the original work on Galileons \cite{Nicolis:2009}). \par 
While we have concentrated on the $1+1$-dimensional case, the symmetry and mechanism for the core results of this note are tied to Carrollian boosts, and more generally supertranslations. Thus, as the core mechanism has been isolated, the arguments presented here generalise to higher dimensions and theories with further symmetries. In short, we have a `no-go theorem'.
\begin{nogo}
One cannot build first-order single (real) scalar field theories minimally coupled to a Carrollian connection on the Carrollian manifolds
$$(M \cong \R^{n+1},~ g = \rmd x^i \otimes \rmd x^j \delta_{ji},~ \kappa = \partial_t)\,,$$ 
that are invariant under Carrollian boosts and admit propagating solutions.
\end{nogo}
In short, intrinsic Carrollian field theories with propagating degrees of freedom must involve additional structure beyond a single minimally coupled first-order scalar field. For instance, following Ecker et al. \cite{Ecker:2024}, we may consider a system with two scalar fields. To sketch the construction, we simply double the jet coordinates $(y, z,  y_t, z_t, \hat{y}_x , \hat{y}_x)$, and note that the  momentum density is given by
\begin{equation}
\mathcal{P} = y_t \, \frac{\partial L}{\partial \hat{y}_x} + z_t \, \frac{\partial L}{\partial \hat{z}_x}\,.
\end{equation}
Now, suppose that the Lagrangian $L$ depends on the first jet coordinates in the particular combination $k := y_t \hat{z}_x - z_t \hat{y}_x$. We then observe that 
$$\frac{\partial L}{\partial \hat{y}_x} = - z_t \, \frac{\partial L}{\partial k}\,, \qquad \frac{\partial L}{\partial \hat{z}_x} =  y_t \, \frac{\partial L}{\partial k}\,.$$
Thus, $\mathcal{P} = - y_t z_t \frac{\partial L}{\partial k} + z_t y_t \frac{\partial L}{\partial k} =0$.  Note this holds exactly and the equations of motion have not been employed (so $\mathcal{P}=0$ both on and off shell). Then, for any section $\phi$, including solutions to the generalised Euler--Lagrange equations, $(j_\mathbb{A}\phi)^* \mathcal{P} =0$. Thus, with the specific coupling of the derivatives of the fields via $k$, the symmetry-imposed condition that the momentum density vanishes identically holds and so does not force the electric/magnetic split. The equations of motion may be pseudo-hyperbolic and so allow propagation. One would still need to check that the action and energy density are well-defined, but in principle, we have a ``swifton'' theory. 
%
%
\section*{Acknowledgements}  
The author thanks Péter Horváthy for his comments on an earlier draft of this note and for pointing out the relevant literature on planar Carrollian dynamics. The author extends his gratitude to the anonymous referees for many helpful suggestions and a careful reading of the manuscript.

%
%

\end{document}